\documentclass[doublecol]{epl2} 

\title{Exploiting B Site Disorder for Phase Control in the Manganites}

\shorttitle{B Site Disorder in the Manganites} 

\author{Kalpataru Pradhan, Anamitra Mukherjee \and Pinaki Majumdar}
\shortauthor{Kalpataru Pradhan, Anamitra Mukherjee and Pinaki Majumdar}

\institute{                    
\inst{} Harish-Chandra  Research Institute,
Chhatnag Road, Jhusi, Allahabad 211019, India\\}


\abstract{
Disorder on the active $d$ element site is usually very disruptive for 
conduction and long range order in perovskite transition metal oxides.  
However, in the background of phase competition such `B site' dopants  
also act to promote one  ordered phase at the expense of another. This 
occurs either through valence change of the transition metal or via 
creation of `defects' in the parent magnetic state. We provide a framework 
for understanding the complex variety of phenomena observed in B site 
doped manganites and identify the key parameters that control the physics.  
Using a spatially resolved analysis of B ions in various manganite phases 
we explain the existing data and predict new situations where highly 
polarisable phase separated states can be created.
}

\begin{document}

\maketitle

Correlated electron systems like the cuprates and  manganites
involve competition between various long range
ordered phases \cite{dag-sc,tok-revs}. 
The interplay of this phase competition with weak disorder 
underlies phenomena like cluster 
coexistence, percolative transport, and colossal response. 
The nature of disorder seems to be crucial for these
effects, as observed in the manganites
\cite{tok-revs}, and   
a low concentration of impurities on the active 
$d$ element site is an effective trigger for phase 
separation 
\cite{b-site-cr-rav,b-site-cr-kim1,b-site-cr-kim2,b-site-cr-mori1,b-site-cr-mori2,b-site-co-mach,b-site-x0,b-site-fe-dop1,b-site-fe-dop2,b-site-mg-dop}
and the associated percolative effects.

The results of such substitution depend  
on the reference state and the chemical nature of the impurity.
For manganites, with rare earth (RE) and alkaline earth (AE)
combination RE$_{1-x}$AE$_x$MnO$_3$, several intriguing results
exist for Mn site  doping. 
$(i)$~Magnetic dopants like Cr 
\cite{b-site-cr-rav,b-site-cr-kim1,b-site-cr-kim2,b-site-cr-mori1,b-site-cr-mori2}, 
Co or Ni \cite{b-site-co-mach}
(but not Fe)  on the Mn site
in a $x=0.5$ charge ordered insulating (CO-I) 
manganite 
promote a percolative ferromagnetic metal (FM-M), while non
magnetic dopants of the same valence do not. 
$(ii)$~The orbital ordered A type antiferromagnet (AF)  at
$x=0$ is
destabilised in favour of a ferromagnetic state \cite{b-site-x0} 
by {\it both} magnetic and non magnetic dopants.  
$(iii)$~In contrast to the cases above, where charge-orbital order
is {\it suppressed}, doping Fe on a ferromagnetic metal
\cite{b-site-fe-dop1,b-site-fe-dop2} 
at $x \sim 0.4$ {\it promotes} a charge ordered insulating state!
On spatial imaging most of these systems reveal phase separation
(PS)
and many of them also exhibit  enormous magnetoresistance. 
It is vital to uncover the organising principle behind this 
diversity of effects,
if we are to exploit B site disorder as a tool for phase
control.

There is unfortunately no microscopic model, let alone
a theory, for 
randomly located B dopants in the manganites.
In this paper we write down the first 
detailed model
for B impurities in a manganite host, and study 
the effect of these 
dopants in a variety of manganite phases
using a real space Monte Carlo technique. 

Our principal results are the following: {\bf (i)}~We
discover that the following hierarchy
of effects  arise in all B doping cases:
(a)~change of the effective valence on the  Mn sites, 
(b)~percolation
of the metallic phase through impurity free regions, and
(c)~`reconstruction' of the background magnetism
and charge order by magnetic dopants.
{\bf (ii)}~By exploring the prominent manganite states, and 
different B dopants, we are able to explain most of the
outstanding experimental results. {\bf (iii)}~We
suggest a new experiment to test out an unexplored 
insulator-metal transition driven by B site disorder. 
{\bf (iv)}~We
demonstrate how B impurity locations determine
the percolation pattern  and may allow  atomic level control
of current paths in a material.

The simplest classification of B site dopants is in terms of
their valence in the manganite host. 
Among the usual dopants  
Zn, Mg, and Co, are divalent,
{\it i.e}, in a $2+$ state, 
Ni, Cr, Fe, Sc, and  Al  are trivalent, while Ru,  Sn, and Ti
are tetravalent.  
Some elements can exist in multiple valence states, 
{\it e.g},  Ni can also be $+2$, and Ru can be $+5$, 
but that will not affect our qualitative arguments.
The valence, $\alpha$,  of the dopant affects 
the effective carrier density on the Mn sites through the 
charge neutrality requirement on the compound
RE$^{3+}_{1-x}$AE$^{2+}_{x}$Mn$^{3+\nu}_{1-\eta}$B$^{\alpha}_{\eta}$O$^{2-}_3$,
where 
$\eta $ is the \% of  B site doping, and we write 
the Mn valence  as  $3 + \nu$. This yields  
$\nu(\eta, \alpha, x) = (x + \eta(3 -\alpha))/(1 -\eta)$. 
The effective $e_g$ electron count on  Mn is $n = 1-\nu$,
modified from $n_0 = 1 -x$ at $\eta=0$. 
This change of effective carrier density can itself 
drive phase change as we will see later.
Secondly, dopants with same valence can have 
different effects depending
on their magnetic character. 
Non magnetic dopants only affect the Mn valence, while those
with partially filled $d$ shells can have magnetic
coupling to  the neighbouring Mn moments. 
Experiments suggest that Cr has  strong AF
coupling  \cite{b-site-cr-ni-coup} 
to the Mn ions, 
Ni couples ferromagnetically \cite{b-site-cr-ni-coup},
while Fe, despite its magnetic $d^5$ 
configuration, couples rather weakly.

Based on the inputs above we construct the following model for
manganites with B site dopants:
$H_{tot}= H_{ref} + H_{imp} + H_{coup}$, with
\begin{eqnarray}
H_{ref}~ & =& \sum_{\langle ij \rangle \sigma}^{\alpha \beta}
t_{\alpha \beta}^{ij}
 c^{\dagger}_{i \alpha \sigma} c^{~}_{j \beta \sigma}
 - J_H\sum_i {\bf S}_i.{\mbox {\boldmath $\sigma$}}_i
+ J\sum_{\langle ij \rangle} {\bf S}_i.{\bf S}_j \cr
&&  ~- \lambda \sum_i {\bf Q}_i.{\mbox {\boldmath $\tau$}}_i
+ {K \over 2} \sum_i {\bf Q}_i^2 \cr
H_{imp}~ &=& V \sum_{n \alpha \sigma} d^{\dagger}_{n \alpha \sigma} 
d^{~}_{n \alpha \sigma} \cr
H_{coup} &=& \sum_{\langle n j \rangle \sigma}^{\alpha \beta}
t_{\alpha \beta}^{n j }
 d^{\dagger}_{n \alpha \sigma} c^{~}_{j \beta \sigma}
+ J'\sum_{\langle n j \rangle} {\bf s}_{n}.{\bf S}_j 
+ V_c \sum_{\langle n j \rangle} q_n q_j
\nonumber
\end{eqnarray}

The reference 
`manganite model' $H_{ref}$ is constructed to reproduce
the correct sequence of phases in the `clean' limit.
It involves the nearest neighbour 
hopping of $e_g$
electrons with amplitude $t^{ij}_{\alpha \beta}$ \cite{kp-prl-2007},
Hund's coupling $J_H$, AF   superexchange $J$ 
between Mn spins, and 
Jahn-Teller (JT) interaction, $\lambda$,
between the electrons and the phonon modes ${\bf Q}_i$.
The stiffness of the JT modes is $K$.
We will  not consider RE-AE cation disorder 
in $H_{ref}$ 
since such `A site' disorder masks the B site effects.
The sites ${\bf R}_i, {\bf R}_j$ and 
operator $c, c^{\dagger}$ refer to Mn locations, and
${\bf S}_i$, {\it etc}, are Mn spins.
The local physics of the B ions is
contained in $H_{imp}$ where  
${\bf R}_n$ refers to the B locations and
the operators $d, d^{\dagger}$ 
refer to the B ion $e_g$ states at an energy $V$ above the
center of the Mn band. 
The sites ${\bf R}_n$ are random, with only the
constraint that two B dopants  are not   
nearest neighbours, to minimise   
electrostatic repulsion.
$H_{coup}$ involves $(i)$~$e_g$ hopping matrix elements $t_{\alpha \beta}$,
which we keep the 
same as between the Mn,  $(ii)$~for magnetic
B ions, a superexchange	coupling $J'$ between the B moment
${\bf s}_n$ and the neighbouring Mn moments, and
$(iii)$~a nearest neighbour Coulomb repulsion $V_c$ between the B dopant and
the neighbouring Mn. The total charge $q_j$ on the Mn ion
is $4 - n_j$, where $n_j$ is the $e_g$ occupancy, and $q_n$ is the (fixed)
B ion valence.
Fig.1 is a schematic, showing the relevant levels on
Mn and B, and the coupling between these atoms.
Earlier attempts at modelling B dopants (or Mn vacancies)
only employed a random onsite potential \cite{b-onsite}.

The overall carrier density is controlled through the 
chemical potential term: 
$-\mu(\sum_{i \alpha \sigma}
c^{\dagger}_{i \alpha \sigma} c^{~}_{i \alpha \sigma} 
+  \sum_{n \alpha \sigma} 
d^{\dagger}_{n \alpha \sigma} d^{~}_{n \alpha \sigma})$.
We use the standard limit $J_H/t \gg 1$, and set $K=1$.
In studying magnetic field effects we
will use a coupling 
$H_{mag} = -{\bf h}.(\sum_i {\bf S}_i + \sum_n {\bf s}_n)$, 
where  ${\bf h} = {\hat z} h$ is the applied field.
We treat all spin and phonon degrees of freedom as classical
\cite{class-ref},
and  measure all energies in units of the Mn-Mn hopping $t$
\cite{kp-prl-2007}. The spins, ${\bf S}_i$, {\it etc}, are treated
as unit vectors, and the magnitude of the spin is absorbed in the
couplings $J$ and $J'$.
% -----------------------------------------------------
\begin{figure}
\centerline{
\includegraphics[width=4.2cm,height=6.5cm,angle=270,clip=true]{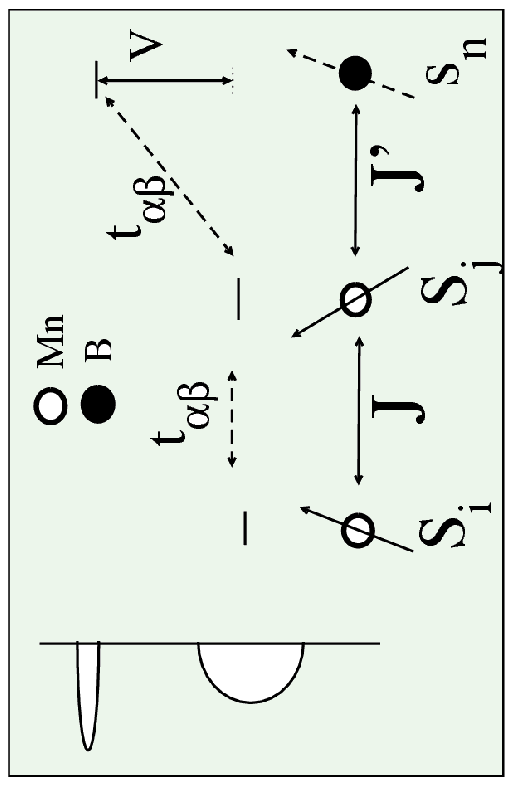}
}
\vspace{.2cm}
\caption{Colour online: 
Energy levels of the B and Mn ions
and the couplings between Mn-Mn and Mn-B.
We show a schematic density of states at the left highlighting
the primarily Mn band and the broadened B level.
}
\end{figure}
% -----------------------------------------------------

The parameter space of the problem is very large. 
We have to contend with
manganite states at different doping $(x)$, different (inverse)
bandwidth  $(\lambda/t)$, and AF strength $(J/t)$. 
The impurities
are characterised by their concentration $(\eta)$, the 
$e_g$ level $(V/t)$, and the magnetic coupling $(J'/t)$.
We use $\lambda/t = 1.6$ as a typical JT coupling,  
$J/t=0.1$ (to capture the
correct phase competition around $x \sim 0.5$),
and $V_c =0.1t$ \cite{coul-ref}.
We will explore several doping levels $x=0.25$, $x=0.4$ and
$x=0.5$ and
study the effect of a low density of B dopants.

We use a Monte Carlo (MC) technique based on the `travelling
cluster approximation' (TCA) \cite{tca-ref}.
% -----------------------------------------------------
\begin{figure}
\centerline{
\includegraphics[width=7.2cm,height=5.5cm,clip=true]{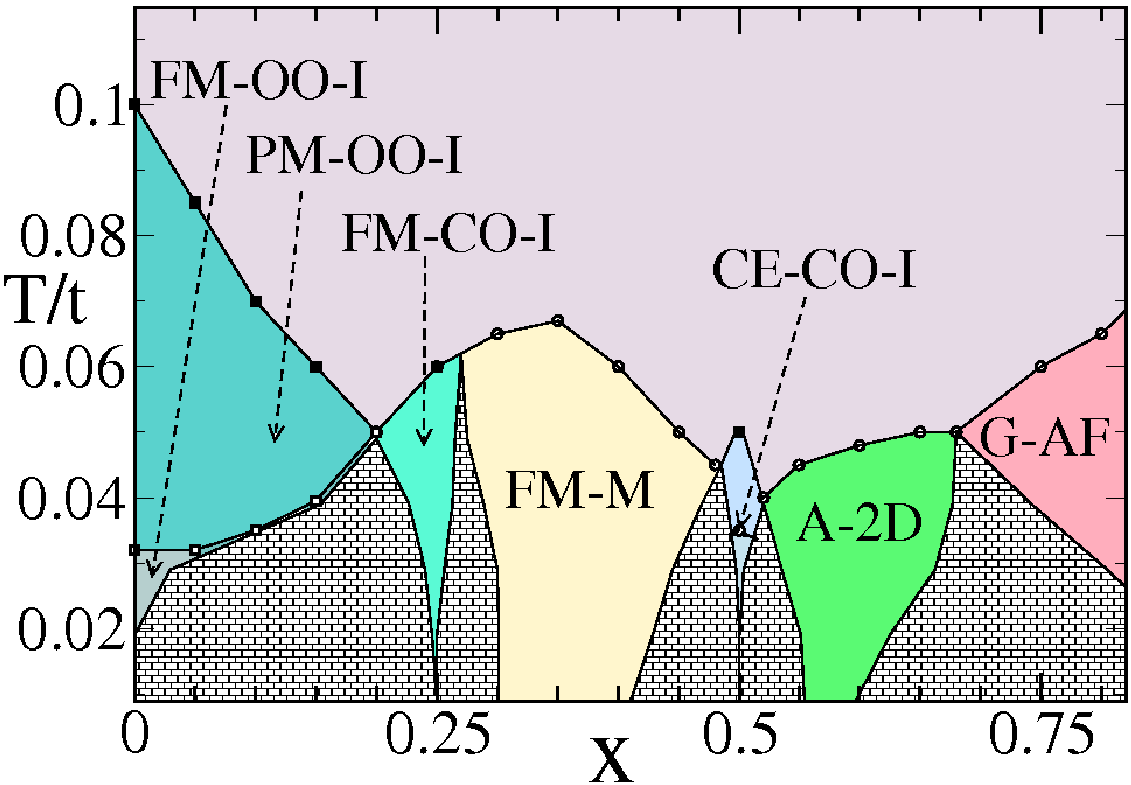}}
\caption{Colour online: The $x-T$ phase diagram of our reference 
model in 2D at $\lambda/t = 1.6$ and $J/t=0.1$. The true 3D transition
temperatures can be estimated roughly as $3/2$ times the 2D scales
indicated here. The phases of interest to us are the ferromagnetic
CO insulator at $x=0.25$, the ferromagnetic metal for 
$x \sim 0.3-0.42$, and the CE-CO phase at $x=0.50$. Shaded
regions indicate phase separation.}
\vspace{.2cm}
\end{figure}
% -----------------------------------------------------
It allows ready access to system size $\sim 40\times 40$ using
a moving cluster of size $8 \times 8$, and handles
disorder accurately. 
The method, and the associated transport calculation, has been
extensively benchmarked by us and used successfully in several
earlier studies \cite{kp-prl-2007,sk-prl-2006}.

The first principle that we wish to illustrate
is phase separation driven by change in effective
carrier density. Fig.2 shows the phases in 
$H_{ref}$ for varying `hole density' $x=1-n$.
The $T=0$ phases, in increasing order of $x$,
are an orbital ordered insulator at $x=0$ \cite{sk-prl-2006}, 
a ferromagnetic charge
ordered insulator (FM-CO-I) at $x=0.25$, a FM-M window between
$x\sim  0.30-0.42$, the CE-CO insulator 
at $x=0.50$, and a magnetic phase (`A-2D'),
with structure factor 
peaks at ${\bf q} = \{0, \pi\},~\{\pi, 0\}$,
between $x \sim 0.55-0.60$. Between these phases are
the shaded windows of PS. If the
carrier density is in one of these PS windows the
system would break up into coexisting patches of the
two adjacent phases. For a system at the  edge of
PS {\it a small valence change driven 
variation in the carrier density
can push it into the PS window. }
Fig.2 shows many such possibilities. These can be exploited by
choosing dopants of suitable valence. 
The tendency towards large scale PS competes with
the fragmenting effect of disorder, leading finally
to  a percolative state.

Let us start with $x=0.25$.
In the clean limit the orbital ordered (OO) JT insulator at
$x=0$  \cite{sk-prl-2006} 
is separated from the  FM-CO-I at $x=0.25$ 
by a wide PS window. The FM-CO-I can be looked upon as 
the charge ordering of doped holes with double the lattice
 periodicity 
in both ${\hat x}$ and ${\hat y}$ directions.
In Fig.2 the phase is bounded on the right also
by a PS window, which separates it from the homogeneous
FM-M.
To  `metallise'  the $x=0.25$ phase
we need impurities that {\it decrease}
the effective electron density, pushing it towards the
FM-M. From our expression for
$n = 1 - \nu$~  at low $\eta$
we obtain $ n \approx (1 -x)  -\eta(3 + x - \alpha) + 
{\cal O}(\eta^2)$, 
so any impurity
with valence $ \le  3$ will serve to lower $n$
and push the system into the $x \sim 0.25-0.30$ PS window. 
The system
could then phase separate, with FM-CO-I and FM-M clusters
whose pattern is controlled by the B ion locations. 

% --------------------------------------------------------------
\begin{figure}
\centerline{
\includegraphics[width=8.0cm,height=8.0cm,clip=true]{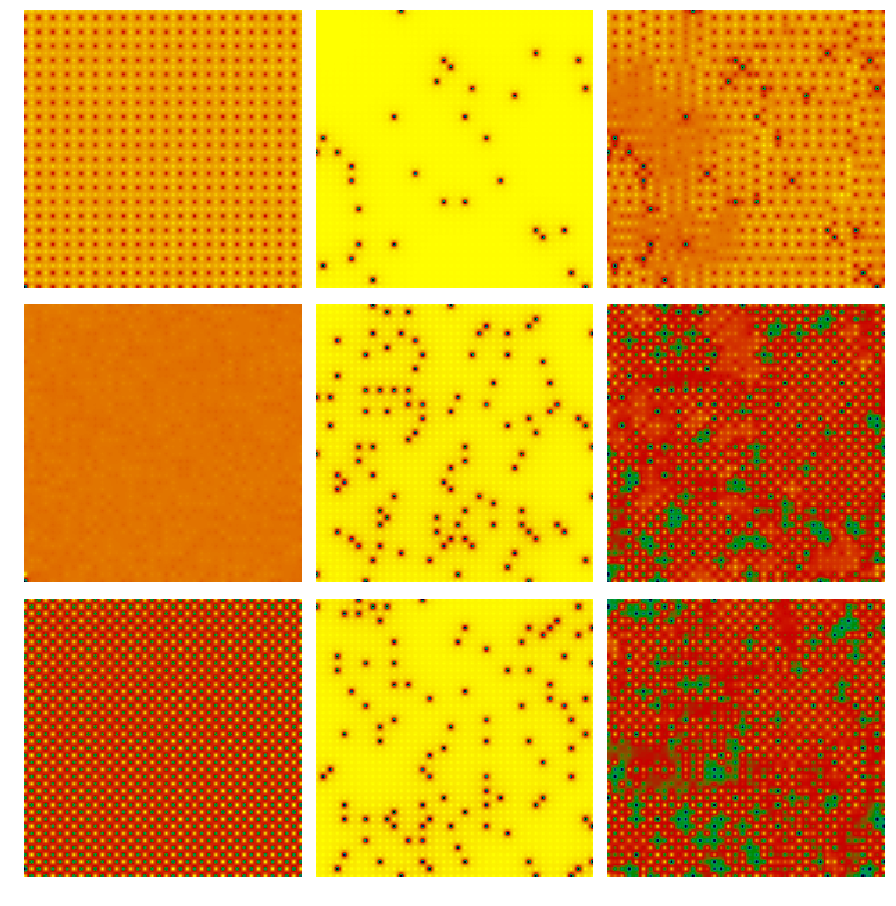}}
\caption{Colour online: The density field $n_{\bf r}$ 
at low temperature.
Columnwise: left - $n_{\bf r}$ in the reference state $(\eta=0)$,
center - location of the B ions, right - $n_{\bf r}$ in the
presence of the B ions.
The top row is for the
$x=0.25$ CO-I state, with  2\% doping of a
non magnetic $2+$  ion with $V=5$.
Middle row is for the
FM-M at $x=0.40$ with  6\% doping of a non magnetic
$2+$  ion with $V=5$. Bottom:
doping on the CE-CO-I at
$x= 0.50$, with a magnetic $3+$ 
ion: $\eta = 6\%$, $V=5$, $J'=0.2$.
 }
\end{figure}
% --------------------------------------------------------------

Fig.3 top row shows a snapshot of 
the charge density field $n_{\bf r}$ in the 
reference CO state at $\eta=0$ (left),
followed by the impurity locations (center), and 
$n_{\bf r}$ (right) in the presence of the B ions.
The impurities are non magnetic $(J'=0)$, 
divalent, with $V=5$
(which is roughly half the Mn bandwidth) and 
$\eta=2\%$. Trivalent impurities yield similar results.
The emergence of homogeneous regions of $n \sim 0.7$ ($x \sim 0.3$)
coexisting
with short range (SR) CO patches is clear. The FM-M live in the
`impurity free' regions, while the SR-CO can coexist with the
impurities at short distance but loses coherence at larger scales.
The percolation of the FM-M patches 
creates a `global' metallic state. 
Fig.4.(a) show the evolution of the
resistivity $\rho(T)$ with $\eta$. Since both
the reference CO phase and the emergent metal are  
FM we cannot get any magnetoresistance in this
regime. It would be interesting to choose $y$ in the 
(La$_{1-y}$Pr$_y$)$_{0.75}$Ca$_{0.25}$MnO$_3$ family
so that the material is at the $T=0$ insulator-metal
phase boundary, and explore the impact of $2+$ or $3+$
B dopants.

Let us shift to $x =0.4$ which is next to the  PS window
separating the FM-M from the $x=0.5$ CE-CO-I phase.
The reference state is a homogeneous metal.
% --------------------------------------------------------------
\begin{figure}
\centerline{
\includegraphics[width=8.0cm,height=8.0cm,clip=true]{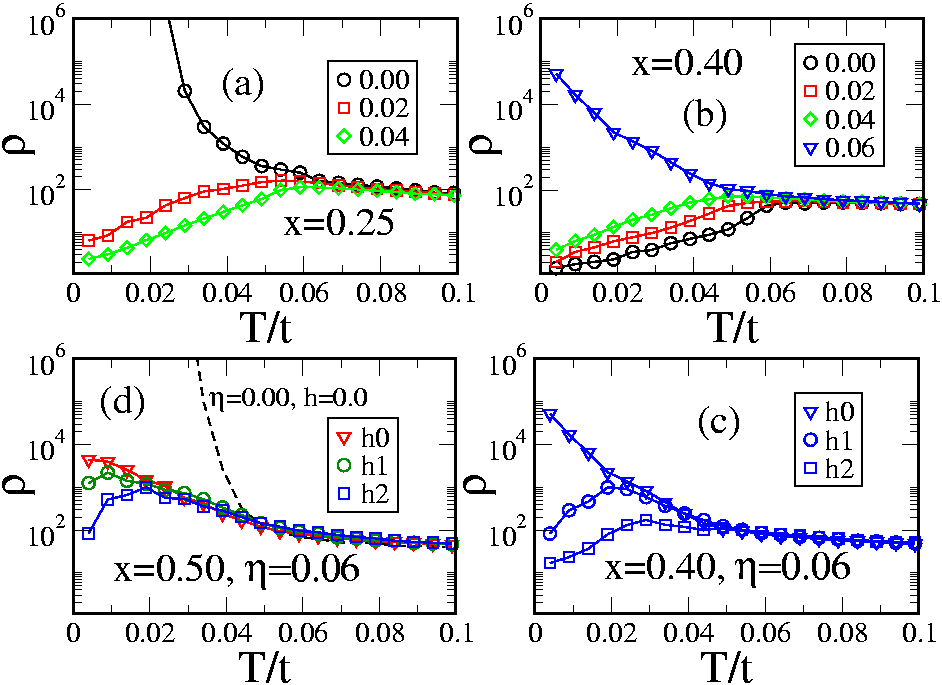}}
\vspace{.2cm}
\caption{Colour online: Temperature and field dependence of
resistivity $\rho(T)$. (a)~Metallisation of the $x=0.25$ FM-CO-I 
with non magnetic dopants of valence $2+$ and $V=5$, 
the curves are for different $\eta$. (b)~Ferromagnetic 
metal to (CO) insulator transition at $x=0.40$ with increasing
concentration of  non magnetic $2+$ dopants with $V=5$, 
(c)~Field response at $x=0.40$ and $\eta= 6 \%$: 
$h_0=0,~h_1=0.002, h_2=0.005$. 
(d)~Resistivity in the undoped CE-CO-I at $x=0.5$
(dotted line),
`metallisation' at $\eta = 6 \%$ with $3+$ dopants $(V=5,J'=0.2)$ 
and field response.
}
\vspace{.2cm}
\end{figure}
% --------------------------------------------------------------
We explore the impact of non magnetic divalent dopants with
$V=5$ 
on this state, and push  $n$ into the PS window.
Such a situation had been explored experimentally early
on by doping Mg \cite{b-site-mg-dop} onto 
Pr$_{0.7}$Ca$_{y}$Sr$_{0.3-y}$MnO$_3$.
A metal-insulator transition 
in the ground state was seen with increasing Mg doping. 
Similar results were observed on doping Fe \cite{b-site-fe-dop1}
(which is a $3+$ dopant with weak magnetic coupling) 
into La$_{1-x}$Ca$_x$MnO$_3$ at $x=0.37$.

The left panel in the middle row in Fig.3 shows the 
`flat' profile of $n_{\bf r}$ in the undoped FM-M state. 
The central panel shows the dopant locations at
$\eta = 6\%$ and the right
panel shows the $n_{\bf r}$ in the presence of dopants.
The final $n_{\bf r}$  shows
local charge order arising out of `disordering' the FM-M!
This has indeed been reported \cite{b-site-fe-dop2} recently.
The FM order in the ground state is significantly
suppressed with increasing $\eta$, and Fig.4.(b) shows
the low $T$ metal-insulator transition driven by B
site doping. Fig.4.(c) shows the dramatic suppression in  
resistivity of this PS system, at $\eta=6\%$,
in response to a magnetic field.
The disorder driven insulating state can be 
readily `metallised' by a modest field.

Just as depleting the electron density 
converts the ferromagnetic metal at $x=0.4$
to a FM-M+AF-CO-I phase
separated state, we can explore the effect of 
{\it increasing} 
the electron density on the CE state at $x=0.5$.
Since  $ n \approx n_0 - \eta(3 + x - \alpha) $
it is obvious that dopants like Ru or Sn, with valence $4+$,
will increase the carrier density to
$ n \approx 0.5  + 0.5\eta $. We explicitly checked that
this leads to suppression of
CE 
%magnetic 
order 
and charge order, enhanced FM correlation, and
at some $\eta_c$ an insulator-metal transition \cite{kp-pm-ce}. 
A simpler version of this was explored earlier \cite{kp-prl-2007}.
%---------------------------------------------------------------------
\begin{figure}
\centerline{
\includegraphics[width=8.25cm,height=5.5cm,clip=true]{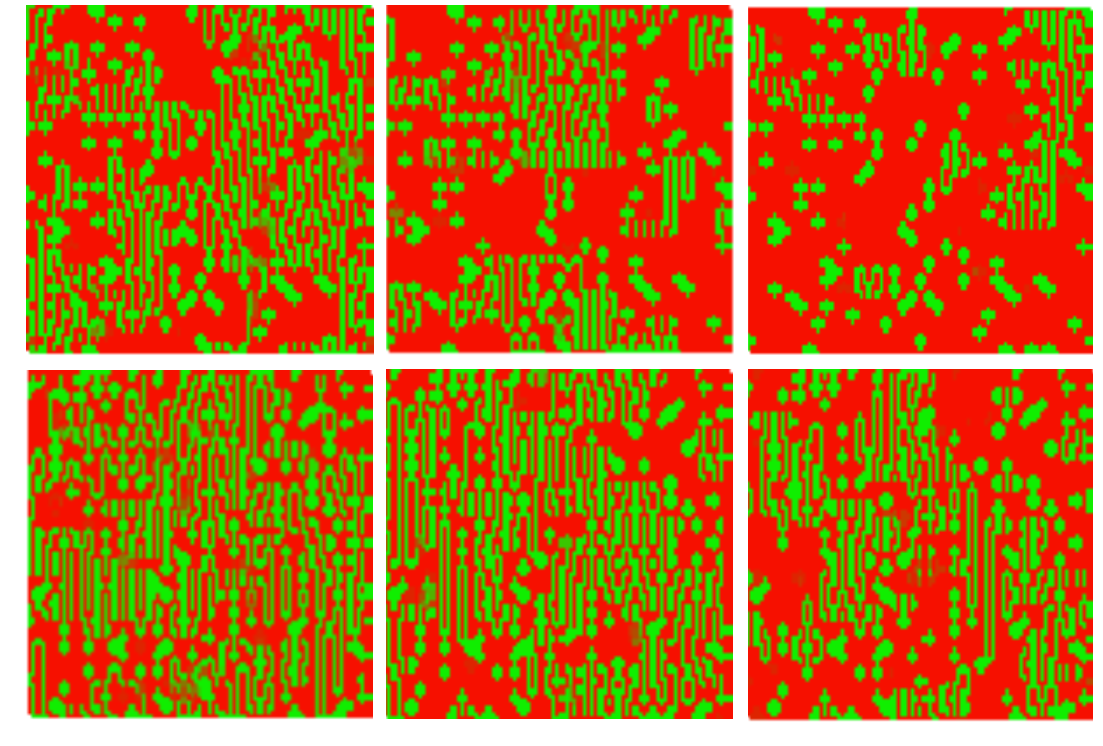}}
\caption{Colour online: Nearest
neighbour  ${\bf S}_i.{\bf S}_j$ from MC snapshots in
the presence of an applied field. Top row, $x=0.4$
with $2+$ dopants at 
$\eta=0.06$, $V=5$, $J'=0$. Bottom row,  $x=0.5$ with
$3+$ dopants at 
$\eta=0.06$, $V=5$, $J'=0.2$. The fields are (left to right)
$h=0,~0.002,~0.005$.
The uniform (red) regions are FM while the stripelike
(green) regions are~AF.
}
\vspace{.2cm}
\end{figure}
%---------------------------------------------------------------------
For B valence $4+$
the magnetic character of the impurity, or Coulomb interaction
$V_c$, are  not relevant.
However, for B ion  valence  
$3+$ the electron density {\it reduces} to 
$n \approx 0.5 - 0.5 \eta$. In that case the carrier density
is not in the PS window between CE-CO-I and FM-M 
but between the CE-CO-I and the A-2D phase!
The resulting state would not be a FM-M 
unless the magnetic aspect of the B dopants 
affect
%had some effect on 
the background antiferromagnetism.

We highlight the case where 
the dopant, {\it e.g}, Cr,  has a strong AF
coupling to the neighbouring Mn. In contrast to the reference
CE phase with $J=0.1$, where a spin has two parallel and two
antiparallel neighbours, using B ions with $J'=0.2$ forces all four
nearest neighbours (NN) to be antiparallel to the B spin.
Due to its low $e_g$ occupancy the B site acts locally like the
`corner site' of a CE chain, and 
the four {\it next NN spins} are also antiparallel to
the B spin. This `$1+8$' cluster \cite{kp-pm-ce}
has a large moment
but at low $\eta$ distant clusters are uncorrelated so 
the magnetic `reconstruction' by itself does not create
a global FM state.
The presence of small 
NN Coulomb repulsion 
$V_c$ suppresses the  $e_g$ charge density
on Mn sites which neighbour the B ions. 
To conserve the electron count, the charge that is pushed out
creates small FM-M regions of locally high density, $n \sim 0.6$,
dominated by the double exchange interaction. These 
FM-M droplets connect the 
otherwise disconnected `$1+8$' clusters.
The resulting pattern is a complex mix of FM-M, FM-CO
and AF regions. 
The ferromagnetism in the magnetic B ion doped case
emerges from a 
combination of 
$(i)$~breakup of the CE pattern by the magnetic dopants, creating
tiny FM clusters, and $(ii)$~their coupling via FM-M droplets 
created by charge pushed out due to the Coulomb interaction.

In addition to converting insulators to metals, and {\it vice-versa}, 
B site doping effects are interesting because the final state is
necessarily inhomogeneous, with possibly huge magnetoresistance (MR).
For example if the PS is between  FM-M and AF-CO-I  
the randomly located B ions fragment the PS state and
the FM-M
domains in such a situation are  weakly linked to
each other at zero field. The large `moments' of the
FM-M domains, see Fig.5, can be aligned by a small 
field leading to enormous increase in conductance.
This large 
polarisability and MR are key signatures of a PS state.
The large changes in resistivity driven by an
applied field are shown in  Fig.4.(c)-(d).  Fig.5 shows
the spatial evolution of the magnetic state for two systems doped
into PS windows. The dramatic
transport response correlates with 
a growth in volume of the
FM-M phase. The effect is stronger for the $x=0.4$ case, but is
also visible for the more complex $x=0.5$ case with $3+$ magnetic
dopants. 

Since we have attempted to model a complex phenomenon, 
and had to make several
approximations to make progress, let us list out some of the
checks \cite{kp-pm-ce} 
we have performed to establish the robustness of our results.
(i)~Any real material will have some degree of A site disorder,
while we have assumed the reference manganite to be
`non disordered'. We have checked that our qualitative results 
survive even if we include binary `A site' disorder of
magnitude upto $0.1$ in our model. In the experiments also
most B site results are on the La-Pr-Ca family, where the
cation disorder $\sigma_A$ is small.
(ii)~Impurity valence fluctuation: in reality it is not 
the valence of the impurity, but its electronic level, that
is fixed. We had assumed B dopants to be of integer valence,
but that is true only if the B site potential $V$ is large, and
the conduction electron density $n_i$ on the B site is $\ll 1$.
We tried out `softer' B potentials, down to $V=1$, and
that leads to $n_i \sim 0.25$ on the B site.
For an assumed valence of 3, say, this implies a modest
change to  2.75, not significantly affecting our argument. 
(iii)~Weak localisation (WL)
effects: we have described `metals' in 2D in the presence of
B site disorder. On large sizes such a system will show
WL effects, and a genuine metallic phase will occur only
in three dimensions.

Let us conclude. 
We have illustrated how one might engineer phase separation
in low cation disorder
manganites by doping B ions of suitable valence 
and magnetic character. The most promising regimes are close to commensurate
filling, $x=0,~0.25,~0.50$, {\it etc}, which are typically ordered
insulating states close to a metal.
Most of the effects can be understood in terms of the 
valence change of  Mn, 
but the B-Mn magnetic coupling is also crucial, particularly
for dopants on the CE phase. The percolative state that emerges 
is typically highly polarisable and has a large low field
magnetoresistance. The current paths in the phase separated
regime are dictated by avoidance of B dopant locations. Since 
presently available techniques allow atomic scale 
manipulation \cite{yazdani} of
dopant locations, B site doping opens up the prospect
of controlling the nanoscale current paths in a 
manganite.

\vspace{.3cm}

We acknowledge use of the Beowulf cluster at HRI.

% -----------------------------------------------------

% --------------------------------------------------------------

\end{document}